# Ultra-Precise, Sub-Picometer Tunable Free Spectral Range in a Parabolic Microresonator Induced by Optical Fiber Bending

**MANUEL CRESPO-BALLESTEROS,**[1,*] **AND MISHA SUMETSKY,**[1]

[1]*Aston Institute of Photonic Technologies (AiPT), Aston University, Birmingham, B4 7ET, UK*

*\**m.crespo@aston.ac.uk*

**Surface Nanoscale Axial Photonic (SNAP) microresonators are fabricated on silica optical fibers, leveraging silica's outstanding material and mechanical properties. These properties allow for precise control over the microresonators' dimension, shape, and mode structure, a key feature for reconfigurable photonic circuits. Such circuits find applications in high-speed communications, optical computing, and optical frequency combs (OFC). However, consistently producing SNAP microresonators with equally spaced eigenmodes has remained challenging. In this study, we introduce a method to induce a SNAP microresonator with parabolic profile. We accomplish this by bending a silica optical fiber in a controlled manner using two linear stages. This approach achieves a uniform free spectral range (FSR) as narrow as 1 pm across more than 45 modes. We further demonstrate that the FSR of the SNAP microresonator can be continuously adjusted over a range nearly as wide as one FSR itself, specifically from 1.09 pm to 1.72 pm, with a precision of ±0.01 pm and high repeatability. Given its compact size and tuning capability, this SNAP microresonator is highly promising for various applications, including the generation of tunable low-repetition-rate OFC and delay lines.**

The quest to develop micro/nano-photonic devices is driven by the aim to replace traditional electronic circuits with their microphotonics counterparts. At the heart of these advancements are optical microresonators, whose tunability represents a fundamental challenge in unlocking their full potential. Traditionally designed for specific applications, optical microresonators lack versatility in their dynamic optical responses. The ability to adjust resonance wavelengths within an optical linewidth enables the creation of reconfigurable optical circuits and facilitates optical modulation, directing a path forward for versatile photonics applications [1–10]. Extending this tunability to cover an entire free spectral range (FSR) or more could revolutionize microphotonics by enabling the alignment of microresonator resonances with various light sources or the integration of multiple microresonators within a single microphotonics circuit. Furthermore, the on-the-fly tunability of microresonators' FSR is essential for applications such as cavity QED [11–13], optomechanics [14,15], generation of tunable optical frequency combs (OFCs) [16–19], frequency conversion [20] and tunable delay lines [21].

Addressing the challenge of tunability, Fabry-Perot microresonators emerge as a viable solution. Their FSR, depending on the material between the mirrors and their separation, can be finely adjusted by physically changing the mirror separation or more complex techniques [22–26]. However, the miniaturization of Fabry-Perot resonators presents its own set of challenges, including complex fabrication processes, the necessity for precise mirror alignment, and sensitivity to environmental factors like temperature changes and mechanical vibrations, all of which can compromise performance.

The tunability of monolithic microresonators, particularly those exhibiting whispering gallery modes (WGM), is also a focus of intense research. They are typically fabricated with crystalline materials in ring, toroidal or spherical shapes that confine light through total internal reflection at their periphery. Their ultra-high Q-factors and superior nonlinear optical coefficients of crystalline materials make these microresonators especially attractive for nonlinear optical processes. However, their FSR (typically in the range of GHz or even THz) is fixed at fabrication, limiting their tunability. Efforts to modify the FSR through piezoelectric actuators or opto-thermal effects have seen limited success due to the large strain or high temperatures required [27–29].

The tunability of WGM microresonators is particularly desirable in the field of OFCs. Since the repetition rate of microresonator-based OFCs are fundamentally determined by the FSR, achieving continuous FSR-tunability is of great importance. To date, alternative solutions involving thermal frequency control have been explored [30,31], but these approaches offer limited tunability or demand the integration of multiple microresonators on a single chip for discrete repetition rate selection [32].

Surface Nanoscale Axial Photonics (SNAP) offers an alternative approach for the WGM microresonator fabrication by nanoscale effective radius variation (ERV) of optical fibers [33–37]. Such a nanoscale variation is sufficient to confine light along the fiber axis, resulting in the formation of axial modes. In a previous work [38], it was shown that bending an optical fiber could form a tunable SNAP microresonator. Its dimensions were adjustable though limited in shape due to the fiber breakage restrictions. A more recent research [39] demonstrated a method to induce SNAP microresonators using two side-coupled, coplanar bent fibers. This method provides control over both the shape and size of the microresonator, as well as the ability to adjust its FSR. However, maintaining a constant FSR across a wide range of eigenwavelengths remains a difficult task in this approach that has yet to be solved.

In this Letter, we address the problem of creation of parabolic (constant FSR) microresonator by bending a silica optical fiber in a controlled manner. Our experimental setup is illustrated in Fig. 1. We attach the ends of the fiber to two plates, forming an arc. The distance between the plates $d$ is adjustable via two linear stages. Such bending not only causes a noticeable geometric deformation, altering the optical path, but also generates stress along the fiber length. The resulting stress causes a change in the refractive index $n$ due to the elasto-optic effect. The combined effect of these two phenomena results in an ERV along the axial length $z$ of the fiber, $\Delta r_{eff}(z)$, that forms our SNAP microresonator. Adjusting the separation $d$ allows us to modify the radius of curvature of the arc, thereby tuning the ERV of the induced microresonator.

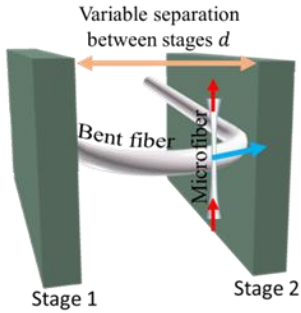

Fig. 1. Illustration of experimental setup. An optical fiber with radius $r_0 = 62.5~\mu m$ is bent with two linear stages, forming a parabolic SNAP microresonator. The bending radius is controlled by changing the distance $d$. The spectrum of the microresonator is scanned using a transverse microfiber connected to an optical spectrum analyzer (OSA).

Light is coupled into the microresonator using a tapered fiber with a micrometer diameter waist, positioned transversely to the bent fiber (see Fig. 1), and connected to an optical spectrum analyzer (OSA) having a 10 MHz (0.08 pm in wavelength) resolution and 60 MHz (~0.5 pm) repeatability. When the wavelength of the input light is close to one of the optical fiber cutoff wavelengths, $\lambda_c(z)$, a WGM is excited. This mode slowly propagates and bounces between the turning points of the induced SNAP microresonator, effectively confining the WGM in the axial direction $z$ and leading to the formation of axial eigenmodes. The slow propagation of the WGMs along the bent fiber is described by the Schrodinger equation [33]. In this equation, the role of potential is played by the cutoff wavelength variation (CWV), expressed as $\lambda_c(z) - \lambda_{c0} = \Delta\lambda_{eff}(z)$, that results from the bending. Here, $\lambda_{c0}$ is the cutoff wavelength of the unbent fiber. The CWV is related to the ERV by the scaling relation $\Delta\lambda_{eff}(z) = \lambda_{c0} \cdot \Delta r_{eff}(z)/r_0$, where $r_0$ is the radius of the fiber, which is consider to remain unchanged.

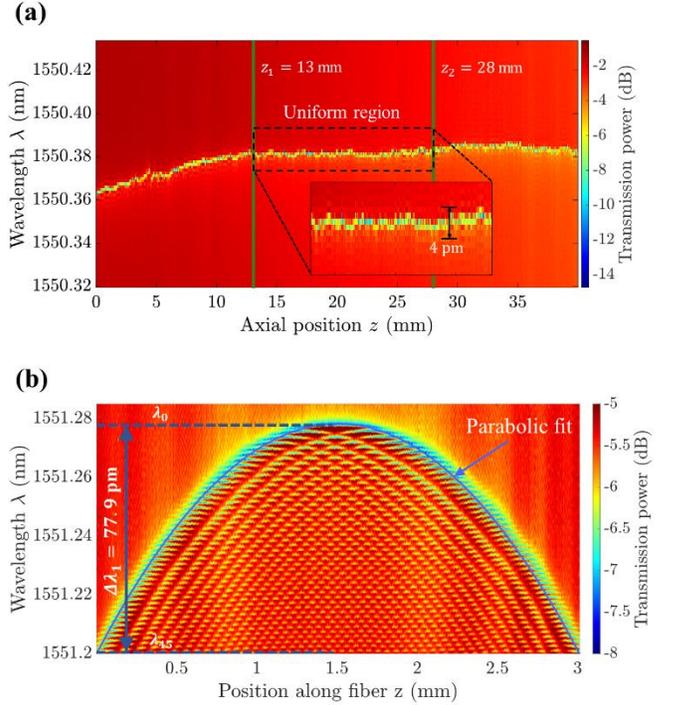

Fig. 2. (a) Spectrogram showing the cutoff wavelength variation (CWV) over a 40 mm length of an unbent optical fiber with $r_0 = 62.5$ μm. A uniform region from $z_1 = 13~mm$ and $z_2 = 28~mm$ has been identified as suitable for bending the fiber to induce a SNAP microresonator. (b) Parabolic SNAP microresonator formed by the bending of the optical fiber. The difference in CWV between the fundamental and the 45th mode is $\Delta\lambda_1 = 77.2$ pm, which corresponds to an ERV of 3.11 nm.

In our experiments, we use a standard silica optical fiber with a radius of $r_0 = 62.5$ μm that has been stripped of its protective polymer coating. We meticulously clean the fiber segment with isopropanol and acetone to eliminate any remnants of the polymer coating. Prior to bending the fiber between the plates, we measure the cutoff wavelength of the unbent fiber along its length $z$. For each position $z$, we measure the transmission power and display it in a spectrogram that plots wavelength against position, as show in Fig. 2(a). We identify a region from $z_1 = 13$ mm to $z_2 = 28$ mm where the cutoff wavelength remains exceptionally uniform, less than 4 pm. This uniformity makes it an ideal region for bending the fiber, as any CWV observed can be confidently attributed to the effect of bending. For a sufficiently small separation distance $d$ corresponding to sufficiently small fiber radius (Fig. 1), the axial resonances of the induced SNAP microresonator were detected. Fig. 2(b) displays the spectrogram of this SNAP microresonator over a 3 mm axial length when $d = 4.35$ mm. Also, in Fig. 2(b), we show a parabolic fit $\Delta\lambda_c(z)$ of the cutoff wavelength of the microresonator's cutoff wavelength $\Delta\lambda_{eff}(z)$:

$$\lambda_c(z) = \lambda_{c0} - \Delta\lambda_0 \cdot \left[\frac{(z-L)^2}{L^2} - 1\right] \qquad (1)$$

Here, $\Delta\lambda_0$ is the maximum CWV over a length of $2L$ of the SNAP microresonator. From the spectrogram in Fig. 2(b), the values of the parameters in Eq. (1) are $\Delta\lambda_0 = 0.0782$ nm, $\lambda_{c0} = 1551.2783$ nm, and $2L = 3$ mm. Impressively, the cutoff wavelength of our SNAP microresonator follows the parabolic fit given by Eq. (1)

without detectable variations. Consequently, the deviations from a constant FSR, $\Delta\lambda_{FSR}$, are dramatically small for at least the first $N = 45$ eigenvalues shown in Fig. 2(b). The spacing between the fundamental eigenwavelength $\lambda_0$ and the 45th eigenwavelength $\lambda_{45}$ is $\Delta\lambda_1 = \lambda_0 - \lambda_{45} \cong 77.2$ pm, yielding the FSR $\Delta\lambda_{FSR} = \frac{\Delta\lambda_1}{N} \cong 1.72$ pm. From the scaling relation $\Delta\nu_{FSR} = \nu_{c0}\Delta\lambda_{FSR}/\lambda_{c0}$ we determine the frequency FSR $\Delta\nu_{FSR} \cong 215$ MHz. Microresonators with such small FSRs are crucial for generating OFCs with low repetition rates (~100 MHz), which are important for applications like high-resolution spectroscopy [40,41] as well as for the fabrication of miniature tunable optical delay lines [42]

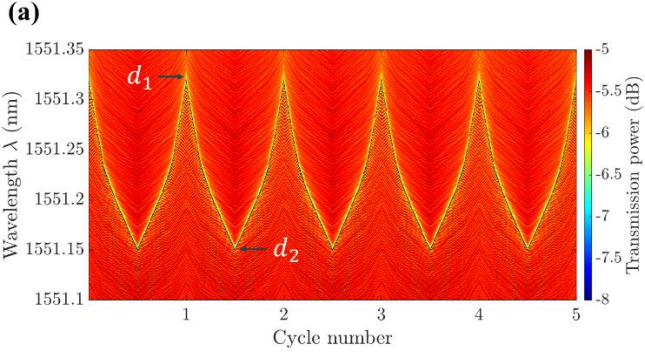

Fig. 3. Repeatability measurement. Cycles of moving the stages between $d_1 = 4.35$ mm and $d_2 = 6.35$ mm.

Next, we demonstrate the tunability of the parabolic SNAP microresonator shown in Fig. 2(b) and repeatability of our system by varying the separation $d$ between the plates from $d_1 = 4.35$ mm to $d_2 = 6.35$ mm (a 2 mm range) in cycles. In each cycle, we incrementally adjust $d$ in steps of 10 $\mu$m, moving from $d_1$ to $d_2$ and then back from $d_2$ to $d_1$. We measure the spectrum of the microresonator at its center for the different values of $d$ during these cycles. Since the input-output microfiber was always in contact with the microresonator, the coupling conditions did not change during the measurement process. The results of this experiment are shown in the spectrogram of Fig. 3. To minimize the effect of temperature variations, which cause global shifts of approximately 1 pm per 0.1°C in the microresonator spectrum, the 5 cycles shown in Fig. 3 were carried out at night. Each cycle lasted 1 hour. This timing was chosen because temperature fluctuations and environmental noise were minimized, achieving a repeatability of better than 1 pm.

From the spectrogram in Fig. 3, we extract the spectrum for each of the 5 cycles at positions $d_1 = 4.35$ mm and $d_2 = 6.35$ mm plotted in Fig. 4(a), and Fig. 4(b), respectively. We observe that the increment in $d$ results in a change of the axial FSR from $\Delta\lambda_{FSR} \cong 1.72$ pm at $d_1$ to $\Delta\lambda_{FSR} \cong 1.09$ pm at $d_2$. We also observe in Fig. 3 and Fig. 4 that the change in $d$ causes a global shift of the microresonator eigenwavelengths. If necessary, this global shift could be corrected by thermo-optical tuning [43].

From Fig. 4, we find that the wavelength separation between the fundamental and the 46th eigenmode, measured at the middle of our microresonator, is $\Delta\lambda_{d_1} = 78.9 \pm 0.4$ pm at $d_1$ and $\Delta\lambda_{d_2} = 50.2 \pm 0.4$ pm at $d_2$. These values fall within the $\pm 0.5$ pm accuracy of our OSA. Using these values and relying on the parabolic nature of the microresonator CWV, we estimate the FSR at $d_1$ and $d_2$ to be $\Delta\lambda_{FSR} = \Delta\lambda_{d_1}/N = 1.72$ pm and $\Delta\lambda_{FSR} = \Delta\lambda_{d_2}/N = 1.09$ pm, respectively, determined with a precision of $\pm 0.4/N = \pm 0.01$ pm.

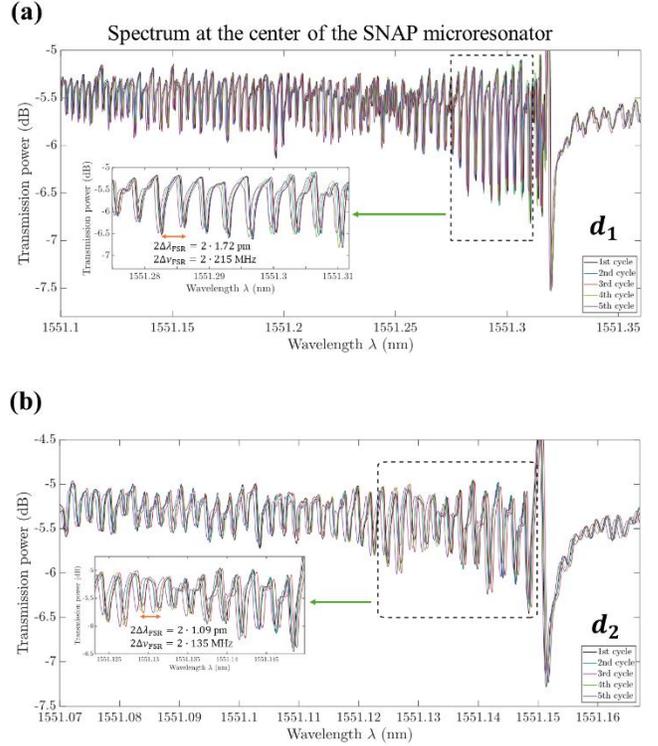

Fig. 4. Spectra measured at the center of the SNAP microresonator for the five cycles displayed in Fig. 3 at positions (a) $d_1$ and (b) $d_2$.

In conclusion, we have demonstrated a method to fabricate a parabolic optical microresonator with tunable FSR, achieved through the controlled bending of a silica optical fiber. The parabolic profile of the CWV is maintained over more than 45 eigenmodes, ensuring a remarkably constant FSR of the order of 1 pm. We demonstrate the ability to continuously tune the microresonator FSR from $\Delta\lambda_{FSR} = 1.72$ pm to $\Delta\lambda_{FSR} = 1.09$ pm with a precision of $\pm 0.01$ pm. We propose that this method can also be used to fine-tune SNAP microresonators fabricated on an optical fiber using other techniques [34–37].

Using fibers with smaller diameters can significantly increase the FSR tunability. Indeed, thinner fibers can withstand tighter bends without breaking. This allows for a smaller radius of curvature of a bent fiber and consequently greater FSR. We also suggest that the bending radius can be made significantly smaller for polymer and soft glass fibers and the investigation of feasibility of forming tunable SNAP resonators in such fibers may be an interesting topic for future research. In our experiments, the Q-factor of the formed SNAP microresonators was limited to approximately $10^6$. This value could potentially be increased to $10^8$ by optimizing environmental and coupling conditions [11]. With this ultra-high Q-factor, we could resolve eigenmodes with wavelength separation of the order of 0.1 pm, 10 times smaller than that obtained in this work.

Overall, the microresonator described in this study stands out due to its compact size and its equally spaced eigenvalue separation that can be tuned with exceptional repeatability. These characteristics make it promising for the generation of OFC with low repetition rates as well as miniature delay lines that can be continuously adjustable. Furthermore, the parabolic profile of the microresonator is required for the optical generation of OFC via parametric modulation [44]. Using the developed method, the formation of combs can be optimized to achieve a broadband OFC spectrum as proposed in Ref. [45]. Future work will explore the possibility of more complex bending shapes to engineer dispersion and expand the bandwidth of generated OFC [46].

**Funding.** This research was supported by Leverhulme Trust grant RPG-2022-014 and by Engineering and Physical Sciences Research Council (EPSRC) grant EP/W002868/1.

**Disclosures.** The authors declare no conflicts of interest.